\documentclass[a4 paper, 12 pt] {article}

\begin{document}
\title{\bf{Monopole-anti-monopole bounded pairs}}
\author{ M. \'{S}lusarczyk\thanks{mslus@alphas.if.uj.edu.pl}
        and A. Wereszczy\'{n}ski\thanks{wereszcz@alphas.if.uj.edu.pl},
       \\ Institute of Physics,
       \\ Jagiellonian University, Reymonta 4, Krak\'{o}w, Poland}
\maketitle
\begin{abstract}
We show that in the dual version of the generalized Dick model
monopole-anti-monopole pairs have finite energy.  It is possible to use
the potential between monopole and anti-monopole to find the mass
spectrum of the glueballs. The results are discussed in connection with
the Faddeev-Niemi model and toroidal soliton solutions. Some others
finite energy configurations are found, both in the magnetic and electric
sector.
\end{abstract}
\section{\bf{ The model}}
Recently it was shown that the Dick model, in the version discussed in
\cite{My}, \cite{My3}, can be
used to model confinement of quarks (this version is slightly different
from the original Dick model \cite{Dick} but there is a sector of the
parameters for which the both models give the same results \cite{My}).
Moreover, the confining potential becomes in agreement with the
phenomenological data
\cite{Motyka}, \cite{Zalewski} for the particular value of the parameter of the model.
In that sense the modified Dick model is a good candidate for the
effective model for the low energy QCD. However, such an effective model
should describe not only quark-antiquark states but also glueballs. In
the present paper a possible way of appearance of the glueballs in the
framework of the modified Dick model is considered. It will occur that it
is possible to look at the glueball states as monopole-anti-monopole
bounded pairs. However, these pairs are found not in the original
modified Dick model but in its "dual" version i. e. in the model where
confinement of the quark sources in the original theory is interchanged
with confinement of magnetic monopoles. Due to that the glueballs and the
scalar mesons appear to be connected by kind of "dual" transformation.
Because of the fact that the interaction of the (non)-Abelian magnetic
monopoles is not understood sufficiently one can find our model
interesting also from the mathematical point of view.
\newline
Let us consider the following action
\begin{equation}
S = \int d^{4}x \left[ -\frac{1}{4} \left( \frac{ \phi }{\Lambda }
\right)^{-8
\delta} F^{a}_{\mu
\nu} F^{a
\mu \nu} + \frac{1}{2} \partial_{\mu} \phi \partial^{\mu} \phi \right],
\label{action}
\end{equation}
where $\delta \geq \frac{1}{4}$ and $\Lambda $ is a dimensional constant.
Indeed one can recognize in this action the dual version of the modified
Dick model
\cite{My}, with the following generalized "dual" transformation
\begin{equation}
F^{a}_{\mu \nu} \rightarrow \phi^{8 \delta} {}^*F^{a}_{\mu \nu}.
\label{dual}
\end{equation}
The dual field tensor is defined in the standard way as $ {}^*F^{a}_{\mu
\nu} =\frac{1}{2} \epsilon^{\mu \nu \rho \sigma } F_{\rho \sigma }$.
The "duality" (\ref{dual}) is equivalent to $\delta \longrightarrow
-\delta$ that means the interchanging of the strong coupling  with weak
coupling sector.
\newline
The fields equations for (\ref{action}) read
\begin{equation}
 D_{\mu} \left[ \left( \frac{ \phi }{ \Lambda } \right)^{-8 \delta}
 F^{a \mu \nu} \right]= j^{a \nu},
\label{f_eq_1}
\end{equation}
\begin{equation}
\partial_{\mu} \partial^{\mu} \phi = 2\delta F^{a}_{\mu \nu} F^{a \mu \nu}
\frac{ \phi^{-8 \delta - 1}}{\Lambda^{-8\delta }},
\label{f_eq_2}
\end{equation}
where $j^{a \mu}$ is the external colour current density.
\section{\bf{Magnetic solutions}}
Let us consider more detailed the Abelian magnetic sector. For example
one can choose
\begin{equation}
A_{\mu }^a=\delta_3^a A_{\mu }, \; \mbox{where} \; \; A_0=0, \; \;
A_i=A_i(x,y,z).
\label{pola}
\end{equation}
Static Abelian monopole solutions can be obtained by means of the
Bogomolny equations. Let us rewrite the energy of a static configuration
as
\begin{equation}
E_N= \int d^3x \left[ \frac{1}{4} \left(\frac{\phi}{\Lambda }
\right)^{-8\delta } F_{ij}F_{ij} +
\frac{1}{2} (\partial_{i} \phi )^2 \right].
\label{enbog}
\end{equation}
The corresponding Bogomolny equation is
\begin{equation}
F_{ij}= \left( \frac{\phi}{\Lambda } \right)^{4 \delta} \epsilon_{ijk}
\partial_k
\phi.
\label{eqbog}
\end{equation}
It is easy to show that this field tensor fulfills the Gauss law
automatically. Moreover, the equation of motion of the scalar field takes
the same form as the Bianchi identity, namely
\begin{equation}
\frac{1}{r^2} (r^2 \phi')' +4\delta \phi^{-1} (\phi')^2 =0.
\label{scalbog}
\end{equation}
One can find the singular  solutions in the following form
\begin{equation}
\phi(r) = \mathcal{A} \Lambda  \left( \frac{1}{\Lambda r} \right)^{\frac{1}{1 + 4 \delta}},
\label{sol1_phi}
\end{equation}
where $\mathcal{A}= [g(1+4\delta )]^{\frac{1}{1+4\delta }}$ and $g$ is
the charge of the monopole. Then the field tensor reads
\begin{equation}
F_{ij}= -g \epsilon_{ijk}
\frac{x^k}{r^3}.
\label{magbog}
\end{equation}
Of course, these Abelian monopoles are the well known Dirac monopoles
with the string attached to them. For these configurations of the fields
we get the energy density
\begin{equation}
\varepsilon = \mathcal{A}^{-8\delta } \frac{g^2}{r^{4}} \left(  \frac{1}{\Lambda r}  \right)^{- \frac{8 \delta}{1 + 4 \delta}}.
\label{energy_dens}
\end{equation}
Indeed, the energy is divergent at large distance that is at $r
\rightarrow
\infty $. One can observe that the energy density has singularity also at $r=0$ but this
singularity is integrable. We conclude that  single monopoles disappear
from the physical spectrum of the theory.
\newline
The dual Dick model has also nonsingular magnetic monopoles labeled by
positive parameter $\beta_0$
\begin{equation}
\phi = \mathcal{A} \Lambda \left( \frac{1}{\Lambda r} + \frac{1}{\beta_{0}} \right)^{\frac{1}{1+4
\delta}}.
\label{sol2_phi}
\end{equation}
The energy density reads
\begin{equation}
\varepsilon = \mathcal{A}^{-8\delta } \frac{g^2}{r^{4}} \left(  \frac{1}{\Lambda r} + \frac{1}{\beta_{0}} \right)^{- \frac{8 \delta}{1 + 4 \delta}}.
\label{energy_dens2}
\end{equation}
These configurations give finite energy
\begin{equation}
E_N = \int \varepsilon r^{2} dr = \Lambda  \frac{4 \delta + 1}{4
\delta - 1} \mathcal{A}^{-8\delta } g^2 \beta_{0}^{\frac{4 \delta - 1}{4
\delta + 1}}.
\label{energy2}
\end{equation}
However, as it was shown in \cite{My3} the finite energy monopole sector
can be removed by adding a potential term for the scalar field. It is
easy to check that this potential can have the following form
\begin{equation}
V(\phi) = \Lambda^4 \left( \frac{\phi }{\Lambda }\right)^{4+8\delta }.
\end{equation}
Vanishing of the monopoles from the phisycal spectrum is not sufficient
to claim that confinement appears. It has to be checked weather a
dipole-like state has a finite energy. We assume that monopole and
anti-monopole lay on the $z$-axis in the distance $\frac{R}{2} $ from the
origin. Then the equations of motion take form:
\begin{equation}
\nabla \cdot \vec{B}=g[ \delta (z-\frac{R}{2}) -\delta
(z+\frac{R}{2})]
\label{q1}
\end{equation}
\begin{equation}
\nabla \times \left[ \left(\frac{\phi}{\Lambda } \right)^{-8\delta}
\vec{B} \right]=0
\label{q2}
\end{equation}
\begin{equation}
-\frac{1}{r^{2}} \left( r^{2} \phi' \right)' - 4 \delta \frac{ \phi^{-8
\delta - 1}}{\Lambda^{-8\delta }} \vec{B}^2= 0.
\label{q3}
\end{equation}
We applied the Ansatze (\ref{pola}) and the magnetic field is given by
\begin{equation}
\vec{B}=\nabla \times \vec{A}.
\label{pota}
\end{equation}
Of course, due to the r.h.s. of (\ref{q1}) the field $\vec{A} $ cannot be
regular everywhere i.e. the Dirac string is still present. In order to
solve the remaining two equations we adopt the procedure presented in the
papers \cite{Adler}, \cite{My2}, \cite{My}. Firstly, we introduce the
cylindrical coordinates $ (\rho,
\phi, z)$, which are natural in our problem. Secondly, we express
the vector gauge potential by means of a scalar flux field $\Phi(\rho,
z)$ in the following way
\begin{equation}
\vec{A}=\frac{\hat{\phi}}{2\pi \rho } \Phi.
\label{anzatz1}
\end{equation}
Using this expression we obtain from (\ref{q2}) and (\ref{q3}) the
equations for the scalar fields
\begin{equation}
\nabla \left[ \frac{1}{\rho} \left(\frac{\phi}{\Lambda}\right)^{-8\delta}
\nabla \Phi \right]=0,
\label{eqdiel}
\end{equation}
\begin{equation}
\nabla^2 \phi +4\delta \left(\frac{\phi}{\Lambda } \right)^{-8 \delta }
\frac{\phi}{\rho} |\nabla \Phi |^2=0.
\label{eqdiell}
\end{equation}
To solve these equations one should find the boundary conditions. This
can be done if we realize that in the monopole-anti-monopole case, the
Dirac string has finite size and connects the monopoles. It is equivalent
to
\begin{equation}
\begin{array}{cc}
 \Phi=0 & \mbox{for} \; \rho =0, \; |z|>\frac{R}{2} \\
 \Phi=g & \mbox{for} \; \rho =0, \; |z|<\frac{R}{2} \\
 \Phi \longrightarrow 0  & \mbox{for} \; \rho^2+z^2 \longrightarrow \infty
\end{array}
\label{bondary}
\end{equation}
where the last condition emerges due to the expectation that energy
density has to fall down to zero at the spatial infinity. It is obvious
that the equations (\ref{eqdiel}), (\ref{eqdiell}) are too complicated to
find analytical solutions. However, it is possible to construct an
approximated solution which obeys the boundary conditions and has finite
energy:
\begin{equation}
\Phi = g\left( \frac{z+\frac{R}{2}}{\sqrt{\rho^2 +(z+\frac{R}{2})^2}} -
\frac{z-\frac{R}{2}}{\sqrt{\rho^2 +(z-\frac{R}{2})^2}} \right),
\label{aprox1}
\end{equation}
\begin{equation}
\phi=\mathcal{A} \Lambda \left( \frac{1}{\Lambda \sqrt{\rho^2 +(z+\frac{R}{2})^2}} \right)^{
\frac{1}{1+4\delta}} -\mathcal{A} \Lambda \left( \frac{1}{\Lambda \sqrt{\rho^2 -(z-\frac{R}{2})^2}}
\right)^{\frac{1}{1+4\delta}}.
\label{aprox2}
\end{equation}
Although these functions do not obey the field equations one can used
them to find the upper bound for the total field energy \cite{My3}. We
get
\begin{equation}
E_{pair}=\beta g^\frac{2}{1+4\delta } \Lambda (\Lambda R)^{\frac{4\delta
-1}{4\delta +1}}.
\label{ener par}
\end{equation}
where $\beta $ is a finite numerical constant. As it was said before, the
quark-anti-quark potential derived from the modified Dick model with
$\delta=3/4$ can be applied to obtain the spectrum of masses in the
quarkonium system. The masses, for many different effective potentials,
were obtained using the Klein-Gordon equation \cite{quarkpot}. It seems
reasonable to expect that the masses of glueballs can be also find in the
similar 'effective potential' model. Now, the effective potential is a
potential between monopole and anti-monopole in the dual version of the
original model (\ref{ener par}). So one can apply the Klein-Gordon
equation and find the masses $M_n $ of the glueball states. Actually, the
effective potential idea in the glueball physics has been recently very
successfully exploited
\cite{Burakovski}. For example, it was shown by West \cite{West} that the lightest
glueball $0^{++}$ can be understood as a bound state of the massless
gluons with the potential:
\begin{equation}
U_{West}=\frac{9}{4} \sigma R - \frac{\alpha }{R},
\label{west}
\end{equation}
where $\sigma $ is the string tension and $\alpha $ is the strong
coupling constant. One can immediately see that our model admits the
confining part of the West potential in the limit $\delta
\longrightarrow
\infty $. This limit is equivalent, in the original modified Dick model, to the
well known string picture of the quarks confinement
\cite{My}, \cite{My3}.
Very similar results were also obtained for the Dual Ginzburg-Landau
model
\cite{Koma}, \cite{Chernodub}.
\newline
\newline
The non-Abelian magnetic monopoles can be obtained if we take into
account all non-Abelian degrees of freedom in the magnetic sector. In
fact, there is the Wu-Yang non-Abalian monopole
\begin{equation}
A^a_i = \epsilon_{aik} \frac{x^k}{r^2}, \; \; \; A_0^a=0,
\label{nieab1}
\end{equation}
\begin{equation}
\phi= \mathcal{C}\Lambda \left( \frac{1}{\Lambda r} +\frac{1}{\beta_0} \right)^{\frac{1}{1+4 \delta}},
\label{new1}
\end{equation}
where $\mathcal{C}=(1+4\delta )^{\frac{1}{1+4\delta }}$. The Dirac string
is no longer present. The point-like singularity which we observe in the
gauge potential is integrable at the energy density level. Moreover,
identical as in the Abelian case, the total energy, for $\beta_0 = \infty
$, is infinite due to the fact that the energy density falls too slowly
in the spatial infinity. One can check that the energy of the non-Abelian
dipole is finite. So, the confining behaviour is visible also in the
non-Abelian theory. The approximated monopole-anti-monopole solution
reads
\begin{eqnarray}
A^a_i & = & \epsilon_{aik}\left( \frac{x^k +x_0^k}{x^2+y^2 +(z+R/2)^2} -
\right. \nonumber \\ & & \left.
\frac{x^k -x_0^k}{x^2+y^2 +(z-R/2)^2} \right)
\label{nieab2}
\end{eqnarray}
Here $x_0=(0,0,\frac{R}{2})$ is the position one of the monopoles. The
scalar fuction has the form (\ref{aprox2}).
\section{\bf{Electric solutions}}
In order to have the general picture of the physics in the dual modified
Dick model we discuss the Coulomb problem. For simplicity we restrict
ourselves to the static non-Abelian source
\begin{equation}
j^{a \mu }= 4\pi q \delta(r) \delta^{3a} \delta^{\mu 0}.
\label{source}
\end{equation}
The field equations take the form:
\begin{equation}
\left[ r^{2} \left( \frac{ \phi}{\Lambda } \right)^{-8 \delta} E \right]^{\prime} = q
\delta(r).
\label{f_eq_5}
\end{equation}
\begin{equation}
\nabla^{2}_{r} \phi = 4 \delta E^{2} \frac{ \phi^{-8 \delta - 1}}{\Lambda^{-8\delta }},
\label{f_eq_6}
\end{equation}
Here $E^{ai}=-F^{a0i}$ and $\vec{E}^{a}=E\delta^{3a} \hat{r}$. The
solutions of these equations form the family parameterized by $\beta_0$:
\begin{equation}
\phi (r) = \mathcal{B} \Lambda \left(  \frac{1}{\Lambda r} + \frac{1}{\beta_{0}} \right)^{\frac{1}{1 - 4 \delta}},
\label{sol3_phi}
\end{equation}
\begin{equation}
E(r) = \mathcal{B}^{8 \delta} \frac{q}{r^{2}} \left(  \frac{1}{\Lambda r}
+
\frac{1}{\beta_{0}} \right)^{ \frac{8 \delta}{1 - 4 \delta}},
\label{sol1_E}
\end{equation}
where $\mathcal{B} = \left[ q (1 - 4 \delta ) \right]^{\frac{1}{1 - 4
\delta}}$. The energy for the family is finite and has the form:
\begin{equation}
E_N = \Lambda \frac{4 \delta - 1}{4 \delta + 1} q^2
\mathcal{B}^{8
\delta} \beta_{0}^{\frac{4 \delta + 1}{4 \delta - 1}}.
\label{energy3}
\end{equation}
Similar to the magnetic sector there is a singular solution of the
Coulomb problem which is divergent at the spatial infinity
\begin{equation}
\phi(r) = \mathcal{B} \Lambda  \left( \frac{1}{\Lambda r} \right)^{\frac{1}{1 - 4 \delta}},
\label{sol4_phi}
\end{equation}
\begin{equation}
E(r) = \mathcal{B}^{8 \delta} q \Lambda^2 \left( \frac{1}{\Lambda r}
\right)^{\frac{2}{1 - 4
\delta}}.
\label{sol2_E}
\end{equation}
However, in that case the singularity is strong enough to remove also
quark-anti-quark solution from physical spectrum i.e. the energy of such
configurations is still infinite.
\section{\bf{Conclusions}}
In our work we have pointed out the model which admits the bounded
monopole-anti-monopole states, whereas the single monopole solution has
infinit energy. Such objects we call magnetic mesons. We assume that
these mesons can be interpreted as glueballs. The potential between
monopole and anti-monopole can be used to find the mass spectrum of the
glueballs. In the limit $\delta
\longrightarrow
\infty $ we reconstruct the confining part of the famous West potential. This result is in
agreement with the standard superconductor picture where the monopole
potential grows linearly
\cite{Chernodub}. Moreover, in our picture, glueballs appear to be "dual" objects to
scalar mesons. That is quark-anti-quark states and glueballs can be
described by means of actions which are connected by very simple "dual"
transformation (\ref{dual}). We believe that this property is not unique
and should be observed in other models which are used to model quarks
confinement. For example, it should be possible to find the
transformation which interchanges the quarks confining sector with the
magnetic monopoles confining sector in the Pagels-Tomboulis effective
model \cite{Pagels},
\cite{My2}.
\newline
Recently it was observed by Faddeev and Niemi \cite{Niemi1}, inspired by
Cho
\cite{Cho1}, that at low energies the appropriate order parameter is an unit
length vector field $n^a$, $a=1,2,3$. For the pure SU(2) Yang-Mills
theory they have proposed the effective action
\begin{equation}
S_{FN}= \int d^4x \left[ m^2 (\partial_{\mu } \vec{n})^2 +\frac{1}{e^2}
(\vec{n}, \partial_{\mu } \vec{n} \times \partial_{\nu } \vec{n})^2
\right],
\label{actionfn}
\end{equation}
where $m$ is a mass parameter and $e$ is a coupling constant. This action
has nontrivial topology. Namely, localized static solutions where
$\vec{n}
\longrightarrow
\vec{n}_0$ for $r \rightarrow \infty$ can be understood as a map
 from $S^3$ to $S^2$.
These maps are divided into homotopy classes $\pi_3 (S^2) \simeq Z$
numbered by the Hopf invariant.  Such knotted solutions were found for
several topological numbers \cite{Niemi2}. One can identify them with so
called magnetic glueballs \cite{Cho} which are supposed to form physical
spectrum of the gauge theory in the low energy limit. Moreover, as it was
mentioned in
\cite{Cho}, the knotted solitons have neither baryonic nor monopoles charges.
Following that we expect that it is possible to interpret knots as bound
states consisting of monopole-anti-monopole pairs. One can suppose that
Faddeev-Niemi model and the dual version of the modified Dick model refer
to the same physics seen from non-topological or topological point of
view respectively. Then the magnetic glueball is a topological object in
Faddeev-Niemi theory, with appropriate Hopf number or a non-topological
soliton in the dual modified Dick model. Unfortunately, we do not know
how the models correspond to each other. However, as it was shown in
\cite{Koma} it is possible to obtain the glueball $0^{++}$ in the toroidal as
well as in the effective potential framework in the dual Ginzburg-Landau
model. Because of that one can try to find the potential representation
of the Faddeev-Niemi model and fit our parameter $\delta $ to it.
\newline
The both problems i.e. the glueballs spectrum as well as the connection
between the Faddeev-Niemi and our model will be considered in the next
papers.
\newline
\newline
We would like to thank Professor H. Arod\'{z} for many helpful comments
and suggestions.

\end{document}